\documentclass[twocolumn]{article}
\usepackage{imr}
\usepackage{graphicx}
\usepackage{amsmath,amsthm,amssymb,amsfonts}

\usepackage[ruled]{algorithm2e} % For algorithms

\newcommand{\calO}{\ensuremath{\mathcal{O}}}

\newcommand{\cf}{\ensuremath{\mathcal{O}}}

\newcommand{\mcal}[1]{\ensuremath{\mathcal{#1}}}
\newcommand{\ie}{\textit{i}.\textit{e}. }
\newcommand{\eg}{\textit{e}.\textit{g}. }
\newcommand{\norm}[1]{\left\lVert#1\right\rVert}
\newcommand{\argmin}{\operatornamewithlimits{argmin}}

\newcommand{\dOm}{\ensuremath{\partial \Omega}}

\newcommand{\vc}[1]{\ensuremath{\mathbf{#1}}}
\def\*#1{\mathbf{#1}} % Bold vectors

\begin{document}

\title{\uppercase{
    Multiple approaches to frame field correction for CAD models
	}}

\author{Maxence Reberol$^{1}\footnote{maxence.reberol@uclouvain.be}$ 
\and Alexandre Chemin$^{2}$
\and Jean-Fran\c{c}ois Remacle$^{3}$}
\date{$^1$Universit\'e catholique de Louvain, Avenue Georges Lemaitre 4, 1348 Louvain-la-Neuve, Belgium}

\abstract{
    Three-dimensional frame fields computed on CAD models often contain
    singular curves that are not compatible with hexahedral meshing.
    In this paper, we show how CAD feature curves can induce non
    meshable 3-5 singular curves and we study four
    different approaches that aims at  correcting the frame field
    topology. All approaches consist in modifying the frame field
    computation, the two first ones consisting in applying internal
    constraints and the two last ones consisting in modifying the
    boundary conditions. Approaches based on internal constraints are
    shown not to be very reliable because of their interactions with 
    other singularities. On the other hand, boundary condition modifications
    are more promising as their impact is very localized. We eventually 
    recommend the 3-5 singular curve \emph{boundary snapping} strategy, which is
    simple to implement and allows to generate topologically correct frame fields.
}


\maketitle

\thispagestyle{empty}
\pagestyle{empty}

\section{Introduction}

In the last decade, frame field based approaches (\S\ref{ss:rel_work}) have become
popular and promising for hexahedral and hex-dominant meshing.
However, state-of-art frame fields are not guaranteed to have the right topology for
full hexahedral meshing \cite{ray2015,viertel2016,Liu_2018}, especially when
the CAD model contains feature curves on the boundary
(\S\ref{s:ff}). 

In \S\ref{ss:ff_analysis}, we show that the 3-5 singular
curves frame field singularity graph issue, i.e. 
singular curve where the hexahedral valence smoothly transitions from three
(index +1/4) on one extremity to five (index -1/4) on the other one, 
is due to the presence of  concave and curved feature curves on the model.
This 3-5 singular curve is the most common problem that arises in
3D frame field topologies. 

Based on these observations, we
investigate four approaches to automatically correct the frame field topology
(\S\ref{s:corr}).  We present the reasoning behind each approach as
well as their pros and cons (frame fields corrected successfully,
failures, and limitations).  Two of the approaches rely on
extrusion of objects inside the frame field: (i) the
extrusion of boundary feature curves (\S\ref{ss:corr_ic}) and (ii) the
extrusion of singular curve extremities (\S\ref{ss:corr_ssl}). 
Extrusion based methods work on simple cases but fail when the extrusion
process is perturbed by other singularities of the frame field.

Another possibility for correcting frame fields is to remove the concave
feature curves, by transforming them into fillets (\S\ref{ss:corr_fillet}). In
other words, irregular 3-5 transitions are much less present in smooth models.
In the smooth case, frame fields are generally compatible
with hex meshing, but the difficulty is to map the singularity graph onto the
non-smooth original geometry. Instead, and it is what we consider to be our
best approach,  we propose to remove the 3-5 singular curves by snapping them
on the boundary and adapting the frame field boundary conditions
(\S\ref{ss:corr_snap}). 

The presence of 3-5 singular curves in frame fields is one of the main issues
that prevents frame-field based hexahedral meshing and automatic blocking. By
removing them, our snapping correction extends the range of application of
frame-field based meshing technology to a larger set of CAD models, where
previous techniques would fail because of incorrect frame field topology.
Nevertheless, this approach is limited to 3-5 singular curves close to the
boundary, which are the most frequent in CAD, and cannot handle arbitrary cases
of 3-5 singular curves.

\section{Related work}
\label{ss:rel_work}

The standard frame field based meshing approach consists in building a smooth
and boundary-aligned frame field (\S\ref{ss:rw_ff_design}) and using it as a
guide to build a hex-dominant
\cite{Baudouin_2014,Bernard_2016,gao2017robust,Sokolov_2016,mindthegap}
or a fully hexahedral mesh 
\cite{Nieser_2011,Li_2012,lyon2016,Liu_2018,Zheng_2018,livesu2019}. When the frame field
singularity graph does not match a feasible hexahedral mesh topology, 
mixed-integer parameterization approaches fail and hex-dominant ones
produce lot of tetrahedra. 
Some heuristic-based attempts to correct the frame fields
have been tried (\S\ref{ss:rw_ff_corr}), but the problem remains
largely unsolved.

\subsection{Frame field design}
\label{ss:rw_ff_design}

In dimension two, crosses are objects made of two orthogonal directions,
invariant by the four rotations of respectively 0, 90, 180 and 270 degrees.
Cross fields are usually represented by 2D vector fields of the form $\vc{f(\vc{x})} =
(\cos(4 \theta(\vc{x})), \sin(4 \theta(\vc{x})))$ \cite{palacios2007}.  To get
a smooth vector field in a 2D domain $\Omega$, the natural
way is to minimize the Dirichlet energy $\int_{\Omega}\norm{\nabla \vc{f}}^2$
under Dirichlet boundary conditions. One issue with this approach is that the
gradient of the frame field tends to infinity at singularities in the
continuous setting.  This is either ignored, as the energy stay finite after
discretization, or addressed by using a scaling scalar field
\cite{knoppel2013}, or more recently by turning to the Ginzburg-Landau theory
\cite{beaufort2017,viertel2019}. In practice, the frame fields obtained with
these different approaches are similar, with singularities whose indices are
compatible with quadrilateral meshing.

In dimension three, frames are made of three orthogonal directions and are
invariant by the twenty-four rotations of the cube. For the generation of boundary aligned
smooth frame field, the ideas are similar to the 2D case: find a convenient
representation of the frame field and minimize the Dirichlet energy. However,
finding a unique and continuous representation is more tricky. The current best
candidates are spherical harmonics \cite{Huang_2011} and fourth-order tensors
\cite{chemin2018}. In both cases, the spaces are of dimension nine and the
frames live on a manifold of dimension three. The Dirichlet energy is minimized
while staying on the frame manifold, which in practice is done either by
optimizing Euler angles associated to frames \cite{Ray_2016} or iteratively
with a non-linear solver and by using recurrent projections on the frame
manifold \cite{chemin2018}. The resulting frame fields exhibit singularities
made of internal curves, that is usually called the singularity graph. As in
dimension two, the norm of the gradient tends to infinity at the singular curves,
which can be seen as extrusions of the singular nodes of a boundary cross field. 

Contrary to the 2D case, frame field singularity graphs do not always
correspond to feasible hexahedral mesh connectivity. This issue has been
extensively described in recent articles \cite{ray2015,viertel2016,Liu_2018}.
To our knowledge, there exist no frame field generation approach that
has any kind of guarantee to provide a singular graph that is
compatible with hex meshing.

\subsection{Frame field correction}
\label{ss:rw_ff_corr}

\begin{figure*}
    \includegraphics[width=1.\textwidth]{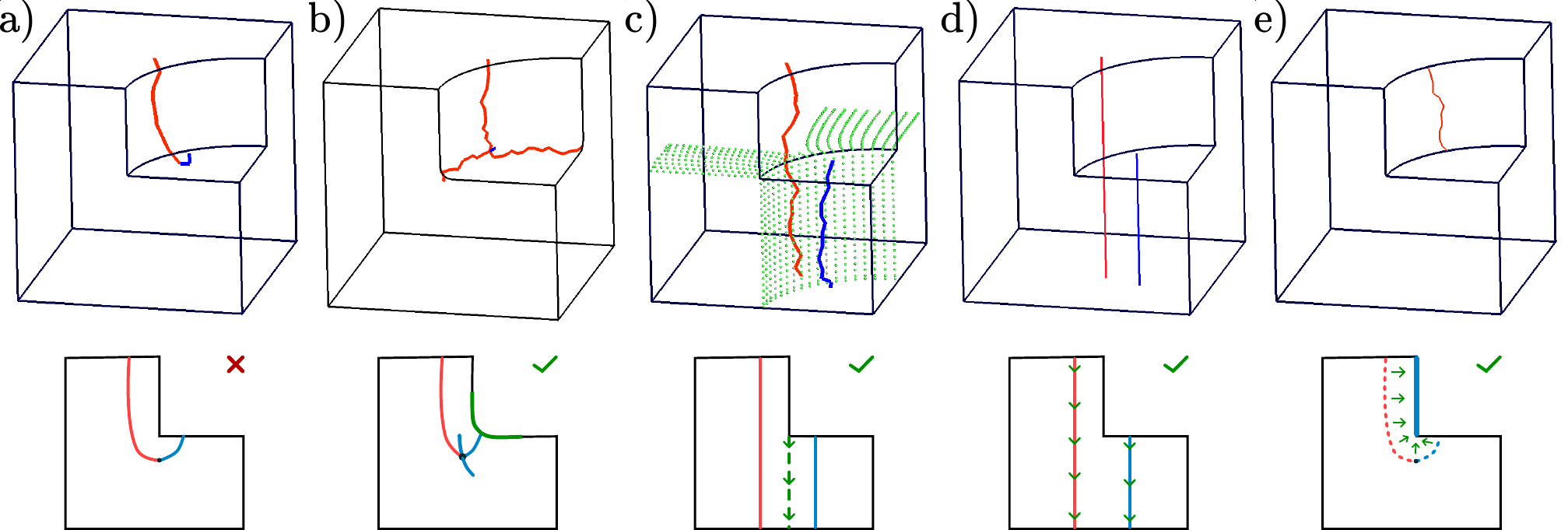}
    \caption{
        a) Natural frame field singularity graph, made of one non-meshable 3-5 singular curve.
        b) When replacing the concave feature curve by a fillet, the new singularity graph
        is made of three valence-five and one valence-three singularity curves, which are
        compatible with hexahedral meshing. 
        c) Via the streamlines (green), the
        feature curve is extruded inside the volume and the resulting singularity graph is
        made of two singular curves (one valence five and one valence three).
        d) Tracing streamlines from boundary singular nodes generates two valid singular curves.
        e) The initial invalid 3-5 singular curve of a) is snapped to the boundary surface, creating
        a new valence three feature curve.
        \emph{Bottom row:} schematic view of the corrections, in the diagonal cut view.
    }
    \label{fig:notch_all}
\end{figure*}

A simple 3D model exhibiting a fundamentally invalid singularity graph is the
\emph{notch model} \cite{ray2015,viertel2016} (Figure \ref{fig:notch_all}), which is the boolean difference
between a box and a cylinder. The singularity graph produced by all existing
frame field methods is made of a single curve whose hexahedral valence is three
on one extremity and five on the second one, that will be called a \emph{3-5 singular curve}
in the rest of the paper. 
In a hexahedral mesh, it is not
possible to have an interior vertex adjacent to only one valence three edge and
one valence five edge (other adjacent edges being regular, \ie valence four). The only valid
vertex configurations are the ones which are topologically equivalent to sphere
triangulations, as described in \cite{Liu_2018}. When a singularity graph contains
a different configuration, \eg a 3-5 singular curve, we say it is invalid or not hex-meshable.

Given a valid singularity graph, there exist methods that allow to compute a smooth frame
field, \eg \cite{Liu_2018,Corman2019}. Thus, correcting a frame field can be
achieved by generating a valid singularity graph, \ie the set of irregular
edges of the associated block decomposition. But generating a valid singularity
graph from scratch remains a totally open problem. An alternative
approach is to start from an initial frame field, possibly
non-meshable, and modify it in order to make it hex-meshable.

In \cite{viertel2016}, a valid frame field for the \emph{notch model} is
built by either manually extruding the concave feature curve inside the
model (producing an internal surface) or by manually adding a fillet to the
feature curve. In the present paper, we propose two approaches (\S\ref{ss:corr_ic}, \S\ref{ss:corr_fillet})
that aim at automatizing these manual interventions.

To avoid the block decomposition degeneracies caused by a 3-5 singular curve,
Zheng and his co-authors \cite{Zheng_2018} propose to replace it by
two singular curves, one with a valence of three and one with a
valence of five, whose geometries are determined by tracing
streamlines starting at the extremities of the 3-5 curve. One of our correction
approach (\S\ref{ss:corr_ssl}) is based on this strategy.

It is also worth mentioning that there can be local defects in the extracted
singularity graph, \eg the \emph{zig-zag} issue. They can be corrected with
local operations, as detailed in \cite{jiang2014}. We are not interested in
these issues in this paper as they are artifacts of discretization choices,
and it is possible to avoid them by using a frame per vertex instead of one per
tetrahedron.

\section{Boundary-aligned smooth frame field}
\label{s:ff}

In this section, we first describe the energy minimization formulation of the boundary-aligned
smooth frame field problem (\S\ref{ss:cf_gen}), which is common to most of the
recent 3D frame field solvers. In this paper, the frame field is discretized on a tetrahedral
mesh $\mathcal{T}$, by defining one frame per vertex (\S\ref{ss:ff_discrete}).
More precisely, we use a continuous piecewise linear approximation of nine
frame coefficients, which fully determine a 4-th order tensor representation of
the frames \cite{chemin2018}.  The following section (\S\ref{ss:ff_analysis}) is a
qualitative analysis of the associated frame field singularity graphs and of
how boundary conditions can lead to configurations which do not correspond to
hexahedral meshes.

As we do not enter into the details, readers unfamiliar with frame fields
should refer to existing articles that explain extensively the theory and
implementation of such fields, \eg \cite{Huang_2011,Ray_2016,solomon2017,chemin2018}.

\subsection{Continuous frame field formulation}

\label{ss:cf_gen}

The goal is to compute a frame field  as smooth as possible and
that is aligned with the boundaries/features of the model.
The natural approach is to translate these requirements into a Dirichlet energy minimization
problem. Minimizing the Dirichlet energy (\ref{eq:dirichlet}) ensures the
smoothness in the domain and the Dirichlet boundary conditions
(\ref{eq:dirichlet_BCs}) enforce the alignment with the model boundary.
Formally, the frame field $\vc{f}$ is the solution to the problem:

\begin{equation} \label{eq:dirichlet}
\vc{f} = \argmin_{\vc{f(\vc{x})} \in \cf}{\int_{\Omega}\norm{\nabla \vc{f}}^2 }
\end{equation}

subject to boundary conditions:

\begin{equation} \label{eq:dirichlet_BCs}
    \begin{cases}
        \vc{f}(\vc{x}) = \vc{g}(\vc{x}) \text{ for } \vc{x} \in \dOm_d \\
        \vc{f}(\vc{x}) \perp \vc{n}(\vc{x}) \text{ for } \vc{x} \in \dOm_s
    \end{cases}
\end{equation}

with $\cf$ the space of frames, $\dOm_s$ a smooth subset of the model boundary
$\dOm$, where we want the frames to be tangent to the boundary, \vc{n} the boundary normal, and $\dOm_d$
another subset of $\dOm$ where we impose the three directions of the frames.
Usually, $\dOm_d$ corresponds to the feature curves of the model, also called
hard-edges or ridges. The symbol $\perp$ means that one of the three
frame directions is parallel to a given vector (tangency constraint).

Before going further, one needs to choose a representation of the frames.  Frames
live in a space, denoted $\cf$, which is the quotient space of the space of rotations
$SO(3)$ with the octahedral group $O$ \cite{solomon2017}. Unfortunately, there
is no simple representation of these objects. To date, two representations have been
proposed: spherical harmonics of degree four \cite{Huang_2011} and fourth-order
tensors \cite{chemin2018}. In both cases, frames live on manifolds of
dimension three immersed in $\mathbb{R}^9$. There is an isomorphism between the two
representations, so they are essentially equivalent.  The
continuity and uniqueness of these representations is a necessary condition to use directly
their nine coefficients to compute distances and gradients, which is analogues
to using the Euclidean distance on the unit circle instead of the circle arc
length.

In the continuous setting, the Dirichlet energy (\ref{eq:dirichlet})
blows up because of the presence of singular curves and this singular
behavior is the cause of many issues for the discretization.

\subsection{Discretization of the frame field problem}
\label{ss:ff_discrete}

To find a numerical solution to the frame field problem introduced previously
(\S\ref{ss:cf_gen}), one needs to choose an approximation space for the frame
field and a numerical scheme to solve the nonlinear problem
(\ref{eq:dirichlet}).

We discretize the frame field on a tetrahedral mesh $\mathcal{T}$ and use a
continuous piecewise linear approximation of the frame coefficients, which can
be coefficients of either the spherical harmonics representation or of the 4-th
order tensor one. The discretized frame field $\*{f}_h$ is now entirely defined
by its coefficients $\*{f}_i \in \mathbb{R}^9$ at each vertex $v_i \in \mathcal{V}$
of the tetrahedral mesh. It is worth noting that inside a tetrahedron, the
linearly interpolated coefficients do not correspond to a frame, but the
closest frame can obtained by projection on the frame manifold $\mathcal{O}$.
Compared to the \emph{one frame per tetrahedron} discretization, the piecewise linear
approximation is more efficient: less unknowns for the same mesh, 
better accuracy \cite{Ray_2016},
possibility to use linear finite elements.

If each $\*{f}_i$ is a frame, then the singularities, which correspond to
infinite gradients in the continuous formulation, cannot be represented at
vertices. With this approach, one possibility is to detect singularities by
looking at axis permutations along edge loops. The smallest loops to look at
are the internal faces of the tetrahedral mesh, which are simply triangles.
The singularity graph is then made of connected chains
of tetrahedra. In practice, it forces the singularity graph to be locally very
distorted as it must be contained inside tetrahedra, whose facets are randomly
oriented in the mesh.

Another way to deal with singularities is to allow the coefficients $\*{f}_i$
to represent objects which are not frames. This is analogous to letting 2D
crosses tend to zero at singularities instead of staying unit vectors. The
advantage is that the singularities are smoother, because they are less
affected by the tetrahedral mesh, but they are also more diffuse and there is
no longer a clear localization of the singularities.

Frame field solvers usually work in two stages : initialization via Laplacian
smoothing of the frame coefficients, without enforcing the frame constraint $f
\in \calO$, followed by a smoothing of the frames, where the frames $\*{f}_i$
must lie on $\calO$ or stay close to it. The process usually converges to a local
minimum which is not far from the initialization \cite{Ray_2016}.

The frame coefficients can be forced to stay on the frame manifold $\calO$ by
either recurring projections \cite{chemin2018} or by optimizing the
Euler angles of the associated rotations \cite{Ray_2016}.

In any case, the Dirichlet energy associated to the frame field tends to
infinity with mesh refinement, but stays finite because of the discretization.
Consequently, these approaches only work on uniform meshes. On non-uniform
meshes, the singularities move to areas of coarse elements as the same
singularity graph topology can be represented while costing less energy.

We use 4th order tensors to represent the frames, and we allow the coefficients
$\*{f}_i$ to deviate from $\calO$ at singularities. This particular choice is
not important for the rest of the paper: the correction techniques we study
can be applied to all energy-minimizing frame field solvers.

\subsection{Frame field singularity graph behavior}
\label{ss:ff_analysis}

Boundary aligned frame fields produced by energy minimizing methods are
interesting because they exhibit singularities that form a graph, which is
topologically similar to the singular edges of a hexahedral mesh.
Unfortunately, as described in the literature  (\S\ref{ss:rw_ff_corr}), this
graph is not always topologically equivalent to a valid hexahedral mesh.

As singular curves cost a lot of energy in (\ref{eq:dirichlet}), an amount
tending to infinity with mesh refinement, the energy-based formulation of 
the frame field problem is asymptotically looking for solutions that minimize the length
of the singular curves. 

In CAD models, there are many feature curves on the boundary where we impose
the frames, via the Dirichlet boundary conditions (\ref{eq:dirichlet_BCs}).
These feature curves split the boundary in multiple patches, which become
independent if we consider the associated surface cross field problems. This
boundary \emph{splitting} leads to the apparition of singular nodes on the
surface, which are necessary to have coherent cross fields that respect
the Poincaré-Hopf theorem.  But from the point
of view of the volume frame field, the boundary singular nodes must be the
extremities of singular curves, as there are no isolated singular nodes in a
3D frame field.

A simple example exhibiting this behavior is a box with a circular arc
imprinted on one face, see Figure~\ref{fig:arcs}.a.. This example can be seen
as a simplification of the \emph{notch} model.
In this example, the two patches on top must have singular nodes to
accommodate their boundary conditions, but the frame field solver, which is
minimizing the Dirichlet energy, do not propagate these singularities inside the
model as this would cost a lot of energy, but merge them as soon as possible in
the volume. Models with such configurations are, for example, the ones containing boolean operations
involving spheres or cylinders that do not go through the whole model, which
are common in CAD modelling.

Our interpretation is that the energy-minimization formulation
(\S\ref{ss:cf_gen}) is similar to the Laplace equation $-\Delta \*f = 0$, even with
the additional constraint $\*{f}(\*x) \in \calO$. Hence, this
is essentially a smoothing kernel that act locally when possible. Constraints
from the boundary that cost energy (non-zero frame gradient) do not propagate
far in the volume. 

On the other hand, hexahedral mesh topological constraints (chords, sheets) propagate on
arbitrary long distances. Thus, hexahedral meshes associated to frame fields
are often different from the ones that a user would produce by manually
building a block decomposition.

An ideal answer to this issue would be a new frame field problem formulation 
that allows better propagation of boundary constraints, but unfortunately none
has been successfully developed up to now.

Another important observation is that even if the frame field have a wrong
topology, one of the frame directions is not affected (the vertical ones in the
\emph{notch} and \emph{box with arc} examples). At the transition from valence
three to valence five in the 3-5 singular curve, the frame field stable
direction along the singularity is no longer tangent to the curve. Exploiting
this stable direction is the basis of the \emph{feature curve extrusion}
(\S\ref{ss:corr_ic}) and \emph{singularity extrusion} (\S\ref{ss:corr_ssl})
correction approaches.  It also worth noting that \cite{viertel2016} tried to
penalize the incoherency between the frame field stable direction and the singular curve
tangents, in order to iteratively correct the frame field, but
it leads to numerical instability and this approach was eventually
unsuccessful.

\section{Frame field correction techniques}
\label{s:corr}

\begin{figure*}
    \includegraphics[width=1.\textwidth]{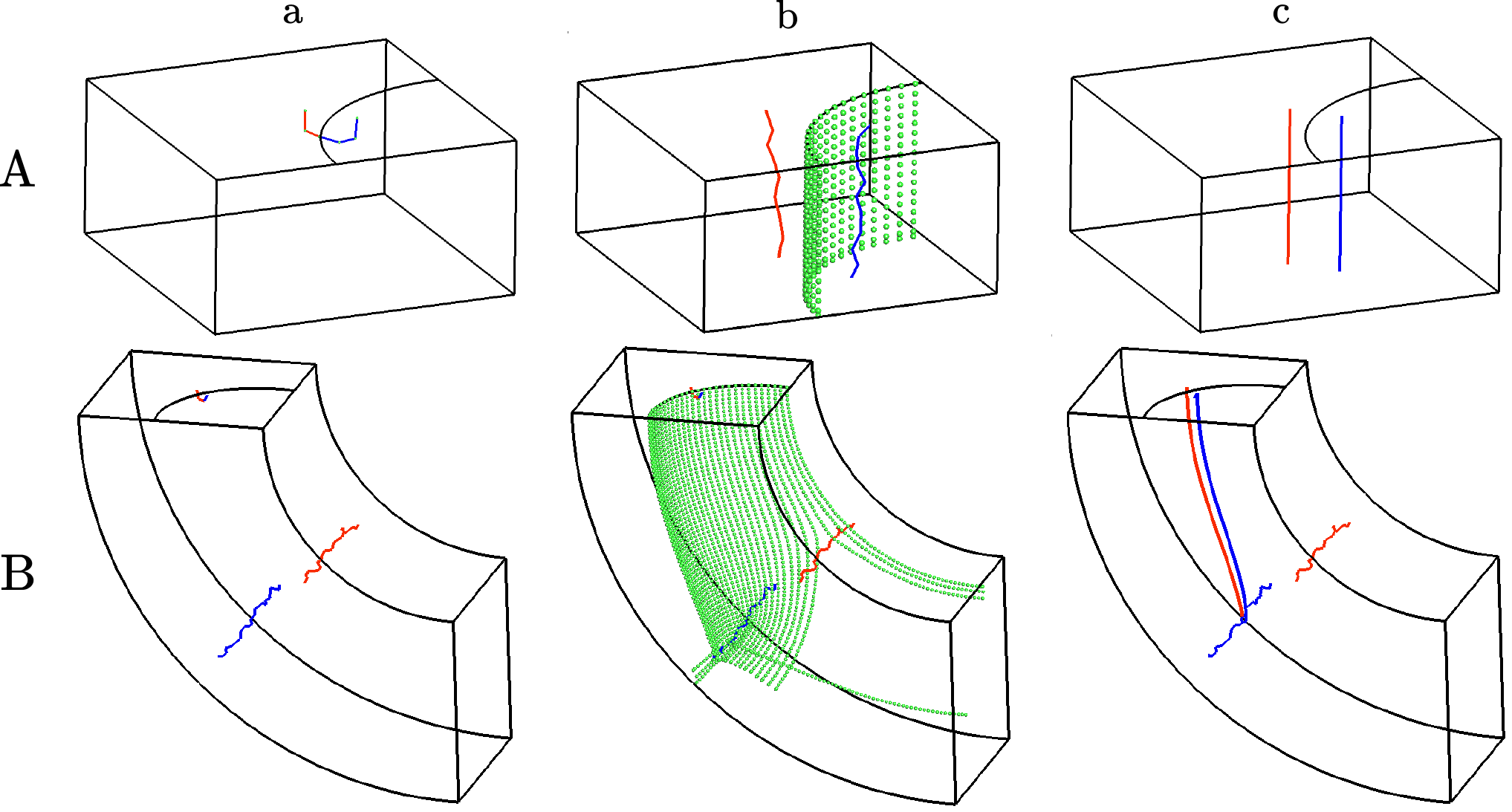} 
    \caption{
        \emph{Model A:} box with imprinted circle arc, simplified version of the \emph{notch model}.
        \emph{Model B:} same model but the box has curvature, which causes two additional singular lines.
        \emph{Col. a:} singularities of the initial frame field.
        \emph{Col. b:} feature curve extrusion, works for A but fails for B because of interaction with
        other singularities, which split the streamline trajectories.
        \emph{Col. c:} singularity extrusion, works for A but fails for B because the streamlines hit
        another singularity.
    }
    \label{fig:arcs}
\end{figure*}

We present and discuss various strategies to automatically correct frame
fields, for them to be suitable for full hexahedral meshing. We focus on
removing the 3-5 singular curves. 

\begin{figure*}
    \includegraphics[width=1.\textwidth]{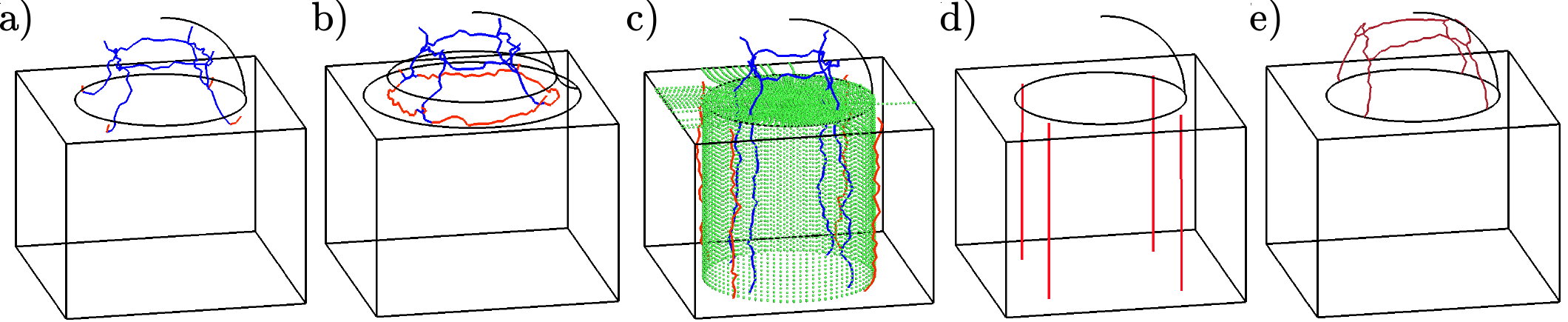} 
    \caption{Half-sphere on top of a box.
        a) Natural frame field singularity graph, containing four non-meshable 3-5 singular curves.
        b) When replacing the concave feature curve by a fillet, the new singularity graph
        contains four additional singular curves of valence five, making the graph
        compatible with hexahedral meshing.
        c) With internal constraints (green), the
        feature curve is extruded inside the volume and the resulting singularity graph is
        valid.
        d) Tracing streamlines from boundary singular nodes replace the four 3-5 singularities by
        four singularities of valence five, but four valence three singular lines are missing. These
        ones should be traced from the internal singular nodes.
        e) The four 3-5 singular curves are snapped to the boundary, forcing snapping of the other singularities.
    }
    \label{fig:hsphere_all}
\end{figure*}

Feature curve extrusion (\S\ref{ss:corr_ic})
and transformation of feature curves into fillets (\S\ref{ss:corr_fillet}) are
automation of the manual corrections applied to the \emph{notch model} in
\cite{viertel2016}. Singularity extrusion (\S\ref{ss:corr_ssl}) has been introduced by
\cite{Zheng_2018} in the context of dual surface construction. Snapping
of 3-5 singular curves to the boundary (\S\ref{ss:corr_snap}) has not been
used previously, to our knowledge.
The four corrections are illustrated side-by-side on Figure~\ref{fig:notch_all} and
Figure~\ref{fig:hsphere_all}.

We show failure cases for the extrusion approaches (\S\ref{ss:corr_ic},
\S\ref{ss:corr_ssl}), where the extruded objects interact badly with other
singular curves (Figure~\ref{fig:arcs}). The transformation of feature curves to fillet
(\S\ref{ss:corr_fillet}) is difficult to apply in practice because it requires
to map the new singularity graph back to the initial geometry, which we
did not automate. The snapping of 3-5 singular curves (\S\ref{ss:corr_snap}) 
seems the most promising approach in our opinion, but it leads to hexahedra with invalid geometry
(zero jacobian at some corners) and a post-processing (insertion of sheets) is required
to have hexahedral meshes suitable for numerical simulation.

\subsection{Feature curve extrusion}
\label{ss:corr_ic}

To recap the previous frame field observations (\S\ref{ss:ff_analysis}): a concave
feature curve can cause 3-5 singular curves, but when it happens there is a stable
direction in the frame field which is not affected.
We propose to use the stable direction of the frame field to extrude the
concave feature curves inside the model.
Once the new internal surface (extrusion of the curve) is generated, we
add internal constraints to the frame field formulation and compute a new
one. While this approach works for some models, it
does not when the extrusion process (streamline tracing) is perturbed by
other singularities.

\paragraph{Tracing streamlines from feature curves}
%% \label{sss:corr_ic_sl}

The simplest approach would be to propagate the feature curves straight into
the model, following the directions of the initial boundary normals. It would
work for simple \emph{blocky} models, but not when the boundaries are
curved, because the new internal surfaces would not follow the curvature of the
model boundaries.

Instead, we extrude the concave feature curves by using certain directions in a
initial frame field. In practice, we achieve this extrusion by tracing
streamlines, that start from the feature curves and end on the first reached
boundaries. In our case, the tracing is done by using a standard fourth order
Runge-Kutta explicit scheme, very similar to the one used with vector fields.
The only difference is that at each new position, we
extract the frame direction which is closest to the previous direction. The reader can
refer to Algorithm~\ref{alg:streamlines} for a more detailed description of the
streamline tracing process.

To determine the initial directions for the propagation, we look at the
appropriate hexahedral valence associated to the feature curve.  It is usually one
or three, and occasionally two for user-inserted curves on smooth surfaces, such
as the circular arc on the Figure~\ref{fig:arcs}, or valence four for
model with cuts whose dihedral angles are close to $360$ degrees. For valence one, there is
no need to propagate the curves as the frame field directions are already
imposed by the surface normals at both sides of the curve. For valence two,
three and four, we respectively extend the curve in one, two and three
directions.

\begin{algorithm}[t]
\SetAlgoNoLine
    \KwIn{
        Initial position $\*p_0$ \\
        \hspace{1.265cm}Initial direction $\*v_0$ \\ 
        \hspace{1.265cm}Step length $h$ for the explicit scheme
    }
    \KwOut{ Streamline $\mathcal{S}$ described by an ordered list of points $(\vc{p}_l)_l$ and
    their associated directions $(\vc{v}_l)_l$}

    $\*{p} = \*p_0 $ \\
    $\*f(\*{p}) = \text{interpolate\_frame\_field\_at}(\*p)$ \\
    $\*v = \text{closest\_direction}(\*v_0,\*f(\*{p}))$ \\
    \Repeat{$\*p$ outside tetrahedral mesh $\mathcal{T}$}{
        append $(\*p,\*v)$ to $\mathcal{S}$ \\
        $\*p = \*p + \frac{h}{6} (\*v_1 + 2 \*v_2  + 2 \*v_3 + \*v_4)$ with \
        $\*v_1 = \text{closest\_direction}(\*v,\*f(\*{p}))$ \
        $\*v_2 = \text{closest\_direction}(\*v_1,\*f(\*{p} + \frac{h}{2} \*v_1 ))$ \ 
        $\*v_3 = \text{closest\_direction}(\*v_2,\*f(\*{p} + \frac{h}{2} \*v_2 ))$ \ 
        $\*v_4 = \text{closest\_direction}(\*v_3,\*f(\*{p} + h \*v_3 ))$ \\
        $\*v = \*v_4$ \
    }
\caption{Streamline tracing in frame field}
\label{alg:streamlines}
\end{algorithm}

Successful applications of the streamline tracing process can be seen on 
Figures \ref{fig:notch_all}.c., \ref{fig:arcs}.A.b and \ref{fig:hsphere_all}.c.,
where the extruded curves are shown with green points.

\paragraph{Internal constraints for frame field}
%%\label{sss:corr_ic_ic}

To compute a new frame field that respects the extruded curve, we add internal
constraints in the boundary conditions (\ref{eq:dirichlet_BCs}) of the frame
field formulation (\S\ref{ss:cf_gen}). The new constraints (\ref{eq:ic}) are
tangential conditions, forcing one direction of the frames, similar to the
boundary conditions on the smooth parts of the boundary.

Consider a streamline $\mcal{S} = \{ (\*p_k,\*v_k) \}_k$, made of 
$(\vc{p}_k)_k$ the points, $(\vc{v}_k)_k$ the associated directions,
started from the edge $e$, of tangential direction $\*t_e$.
Then the associated internal conditions are:

\begin{equation} \label{eq:ic}
    \forall k, \quad \vc{f}(\vc{p_k}) \perp (\*t_e \times \*v_k)
\end{equation}

To apply these internal conditions precisely, we compute a new
tetrahedral mesh containing the points $(\vc{p_k})_k$ 
associated to all the streamlines traced from the concave feature curves.

In our examples, the singularity graphs of the new frame fields are displayed
on Figures \ref{fig:notch_all}.c., \ref{fig:arcs}.A.b and
\ref{fig:hsphere_all}.c.. They correspond to valid hexahedral meshes, similar
to the ones a user would generate by manually building the block decompositions.

\paragraph{Failure cases and limitations}
%% \label{sss:corr_ic_fail}

Extruding a feature curve inside a frame field is equivalent to tracing a
sheet.  If the sheet encounters another singularity orthogonally, then it is
sheared in multiple parts. When the multiple parts can stay inside a future
valid sheet of the hexahedral mesh, as for the horizontal sheet in Figure
\ref{fig:notch_all}.c., then our approach still works.  But if the sheet is
sheared into parts that are sent in different arbitrary directions, then the
new internal surface makes no sense and should not be used to constrain the
frame field, as it would definitely produce an incoherent frame field. This
failure case is illustrated on Figure \ref{fig:arcs}.B.b..

The feature curve extrusion is also not applicable to models 
where the frame field contains limit cycles, such as the Nautilus example in 
Figure 1 of \cite{viertel2016}. In such cases, which are rare, the streamlines
will spiral indefinitely without reaching any boundary. They can be detected
by monitoring the streamline lengths.

Regarding the model on Figure \ref{fig:arcs}.B., other frame field solvers
could produce a frame field without the pair of valid singularities that accommodates
the curvature. In that case, the feature curve extrusion correction would work.
However, it is still possible to add features to the model, \eg boolean
difference with a cylinder, that would introduce singularities to perturb the
extrusion of the feature curve.

The initial 3-5 singular curve is a local solution of the frame field solver to
the boundary constraints. By extruding the feature curve using the stable
directions, we are trying to generate global constraints, as they typically 
propagate through the whole model. We think this approach cannot work in the
general case, as there is no reason that the feature curve conserves its shape
during the extrusion process in the frame field.

\subsection{Boundary singular node extrusion}
\label{ss:corr_ssl}

Another possible extrusion approach is to propagate the boundary singular nodes
via the frame field stable directions, instead of propagating the concave
feature curves. This approach has been implemented in \cite{Zheng_2018} to
separate the contour of dual sheets, in the context of block decomposition.
The additional step in our case is to compute a new frame field that respect
these forced singularities.

The hope with this approach, compared to the feature curve extrusion
(\S\ref{ss:corr_ic}), is that node extrusion produce curves that would not
interact with other singularities, as they are much smaller than surfaces.

To extrude the singular node extremities of 3-5 singular curves, we use the
same streamline tracing method (Algorithm \ref{alg:streamlines}) that was used
for curve extrusion (\S\ref{ss:corr_ic}).
For each extremity, we trace a streamline in the direction of the stable direction,
until it reached a boundary.

The streamline vertices are added in a new tetrahedral mesh and specific
internal conditions are applied on them: forcing them be singular frames with a
unique imposed direction.  Another
alternative could be to use frame field generation methods with fixed
singularity graph \cite{Liu_2018,Corman2019}, but it would require imposing
an entire singularity graph and we prefer to use a more flexible frame field
solver that still has the possibility to displace the previous valid singularities, and
potentially to spawn new ones.

Successful applications are shown on Figures \ref{fig:notch_all}.d.,
\ref{fig:arcs}.A.c and \ref{fig:hsphere_all}.d.. They correspond to the same
valid hexahedral meshes as the ones generated by the previous approach
(\S\ref{ss:corr_ic}). On Figure \ref{fig:hsphere_all}.d., the four internal
valence three singularities are missing because we did not implement the
tracing from an extremity which is an internal node, but it would work.

\paragraph{Limitations} This approach suffers from the same failure case 
than the feature curve extrusion, which is shown on Figure \ref{fig:arcs}.B.c..
Both streamlines hit an existing singularity during the tracing process, leading
to an incoherent singularity graph.

The singularity extrusion correction technique may have a better success rate
than the feature curve extrusion on specific models (less risk of collision),
but it suffers from the same fundamental flaw (interaction with other
singularities) and does not work in the general case.

\begin{figure*}
    \includegraphics[width=1.\textwidth]{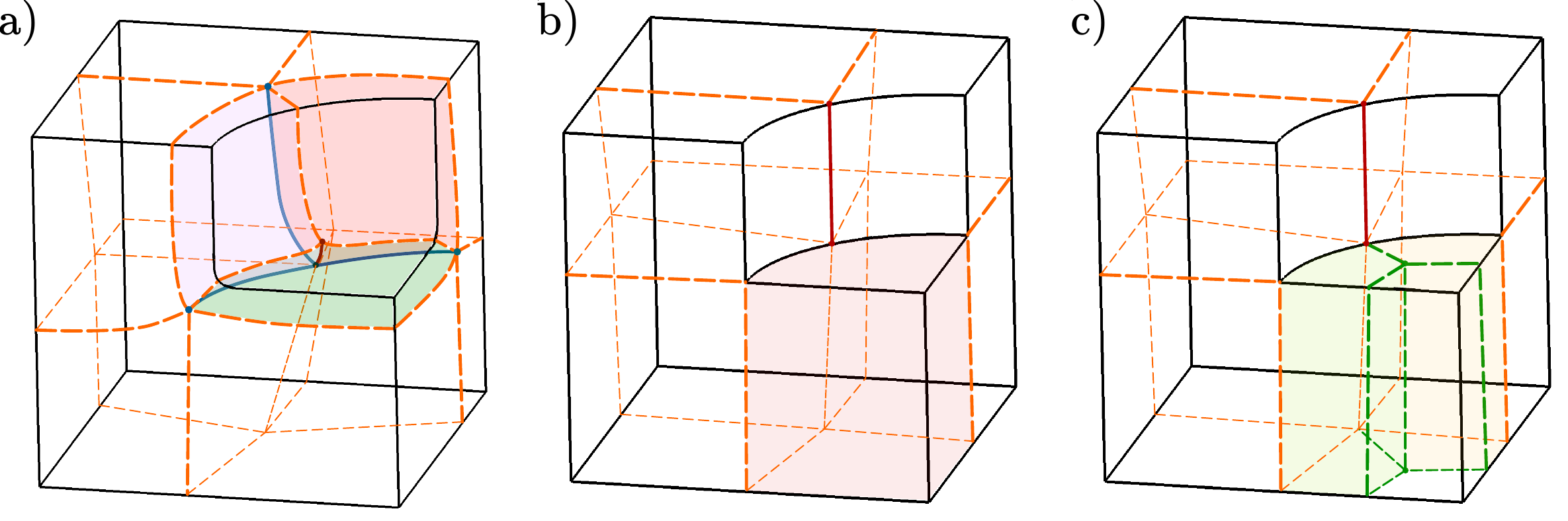} 
    \caption{
        Block decomposition of the \emph{notch} model after correction. 
        a) Feature curve transformed to a fillet. The block decomposition is valid.
        b) 3-5 singular curve snapped to boundary. The block decomposition is
        topologically valid but the colored block has a flat corner.
        c) Refinement of the block with a flat corner. The decomposition is 
        topologically and geometrically valid.
    }
    \label{fig:notch_blocks}
\end{figure*}

\subsection{Transforming concave feature curves into fillets}
\label{ss:corr_fillet}

In the frame field analysis section (\S\ref{ss:ff_analysis}), we explained that
non hex-meshable 3-5 singular curves are caused by feature curves, at least in the
cases we studied. We can try to remove all hard-edges and compute frame
fields on smooth models, assuming that energy-minimizing frame fields
on smooth models are less likely to contain 3-5 singular curves. 

\begin{figure*}
    \includegraphics[width=1.\textwidth]{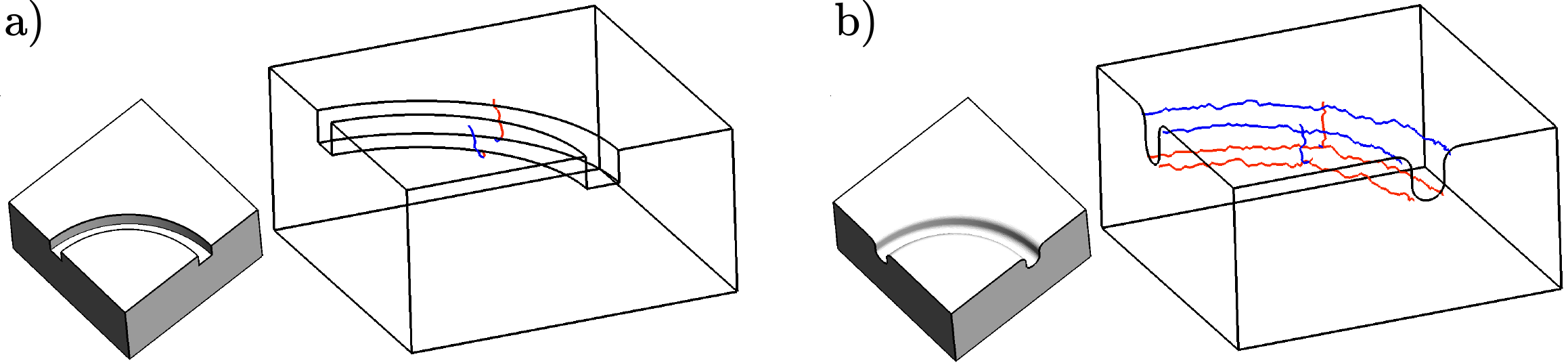} 
    \caption{
        a) Box with a circular rectangular groove on the top. The singularity graph is made
        of two 3-5 singular curves.
        b) The feature curves of the rectangular groove are transformed into
        fillets, making the geometry smooth. This change introduces singular
        curves along the groove, that connect with the previously invalid 3-5
        singular curves. The new singularity graph, which contains four
        internal nodes, is compatible with hex meshing.
    }
    \label{fig:groove}
\end{figure*}

Even if smooth models can contain invalid 3-5 singular curves, \eg the
\emph{rockerarm} in Figure 10 of \cite{Liu_2018}, they are much less widespread
than with CAD models in our experience.  We think it is worth trying to exploit
the smoothing of feature curves to correct frame fields.
A feature curve (made of hard-edges) can be replaced by a smooth transition
between the two surface patches adjacent to the curve.  In terms of CAD
modeling, this is equivalent to placing a fillet on the curve, as illustrated
on Figure \ref{fig:notch_all}.b., Figure \ref{fig:hsphere_all}.b.  and Figure
\ref{fig:groove}.b..

Smoothing a convex feature curve (hex-valence of one) induces an internal
singular curve of valence three in the frame field and smoothing a concave
feature curve (hex-valence of three) induces an internal singular curve of
valence five. If the feature curve that is replaced by a fillet was responsible
for a 3-5 singular curve, the newly introduced singular curves join the
invalid curve at its transition between valence three and five, creating a
hex-meshable singular node. In the case of the \emph{notch} model 
(Figure \ref{fig:notch_all}.a. and b.),
the new singular node connects
one valence-three and
three valence-five singular curves after the insertion of the fillet. The associated
block decomposition is shown on Figure \ref{fig:notch_blocks}.a..
Another application is illustrated on Figure \ref{fig:hsphere_all}.b., where the frame
field of the smoothed model contains four additional valence-five singular curves,
making it suitable for full hexahedral meshing.

From the point of view of the block decomposition, the \emph{fillet} correction
introduces a new hexahedral layer (the colored blocks on Figure
\ref{fig:notch_blocks}.a.). This layer was partially present in the initial
frame field, which contained the 3-5 singular curve, but it was pinched and it did
not correspond to topological blocks (six quadrangular faces). The smoothing of
the hard-edges allows to recover valid blocks. 

Another example of singularity graph obtained after transforming the feature
curves into fillet is illustrated on Figure \ref{fig:groove}. This model is a
volume version of the \emph{box with arc} model, but with the arc replaced by a
groove. As with the \emph{notch} model, the smoothing of the feature curves introduces
new singular curves that connect with the 3-5 singular curves, making all the singular nodes
hex-meshable.

This approach is interesting because it is a local approach, which implies
local modifications of the frame field, well in accordance with the spirit of
the energy-minimizing frame field formulation. 

\paragraph{Limitations}
Unfortunately, it is not
straightforward to implement it in an automatic way.
Once a hex-meshable frame field, or equivalently its singularity graph, is
computed on the model with fillets, it must be brought back to the initial
model with hard-edges. Before trying to develop an automatic technique for
building such mapping, it is interesting to do it manually and observe the
result.

By carefully looking at the block decomposition on the smoothed \emph{notch}
model on Figure \ref{fig:notch_blocks}.a., we can see that the valence-three
boundary singular node has been positioned at the center of the fillet, where
the curvature is maximal. In the initial model, this node has to be mapped on
the center the concave feature curve. This implies that the topological block
in green will have two adjacent edges on this curve, forming a flat angle. This
geometry is not valid from a numerical analysis point of view, as the jacobian
of the hexahedra would be zero on this corner.

Instead of pursuing this approach, which would require lot of engineering, we
focus on a similar but simpler one: the \emph{boundary snapping} of 3-5 singular curves
(\S\ref{ss:corr_snap}).

\subsection{Boundary snapping of 3-5 singular curves}
\label{ss:corr_snap}

The previous \emph{fillet} approach (\S\ref{ss:corr_fillet}) corrects the
frame field singularity graph by adding singular curves in order to make all
junctions hex-meshable.
Another way to deal with the 3-5 singular curves is to make them disappear,
instead of enriching them.  By
observing the block decomposition associated to the \emph{fillet} correction
on Figure \ref{fig:notch_blocks}.a., we can see that there is a layer of blocks
close to the fillet (the colored ones). If we remove this layer, we
get a new block decomposition where the singular curve is on the boundary, see
Figure \ref{fig:notch_blocks}.b.. We propose to mimic this process by directly
snapping 3-5 singular curves to the boundary, skipping the \emph{fillet}
correction (Figure \ref{fig:notch_blocks}.b.).
A drawback is the resulting block decomposition has blocks with zero jacobian
at some corners, but it was already the case with the \emph{fillet}
correction, and the final geometry can be improved by refining some blocks
in a post-processing phase (Figure \ref{fig:notch_blocks}.c.).

\paragraph{Singularity snapping}
Our snapping strategy is simple. For each 3-5 singular curve, we snap
both extremities. If an extremity is a boundary singular node, it is snapped to
the closest feature curve, if it is an internal singular node, it is snapped to
the closest point of the boundary surface.
Other singularities (initially valid) may have only one of
their extremity snapped, if so the other is also snapped. This process is applied
iteratively until all necessary singularity extremities have been snapped.

Once the extremities are snapped, the geometry of the curve on the boundary must
be recomputed. A simple approach is to take the shortest path between both
extremities on the boundary triangle edges, \eg Figure \ref{fig:notch_all}.e. and
Figure \ref{fig:hsphere_all}.e., but a more accurate one is to build a new
triangular mesh of the boundary that includes smooth curves joining the snapped
extremities.

\paragraph{Corrected frame field} To generate a frame field that respect the
snapped curves, the boundary conditions (\ref{eq:dirichlet_BCs}) of the frame
field formulation (\S\ref{ss:cf_gen}) must be adapted. Close to the snapped
curves, which can be seen as new feature curves, the corrected frame field
should not be aligned with the boundary normals.  From the hexahedral mesh
point of view, we would like to have edges of valence one or of valence three
on parts of the surface boundary that are smooth, \eg locally flat.

On the snapped curves, instead of imposing tangency to the boundary normal, we
impose frames (Dirichlet boundary conditions) that are tangent to the curve and
45-degrees rotated from the boundary normal (along the curve axis).

Feature curves that received a snapped extremity are split and new boundary
conditions are obtained by linearly interpolating the frames at both extremities of the
split curves.

To avoid incompatible boundary conditions, we also remove the boundary
alignment constraints on the vertices close to the snapped curves. These frames
become free, allowing a smooth transition from the frames of the snapped curves
to the frames on the rest of the model boundary.

While the new frame field is no longer boundary aligned everywhere, the affected
areas remain localized and the resulting frame field is still similar to the
initial one, minus the 3-5 singular curves that have been snapped.

\paragraph{Applications}

\begin{figure*}
    \includegraphics[width=1.\textwidth]{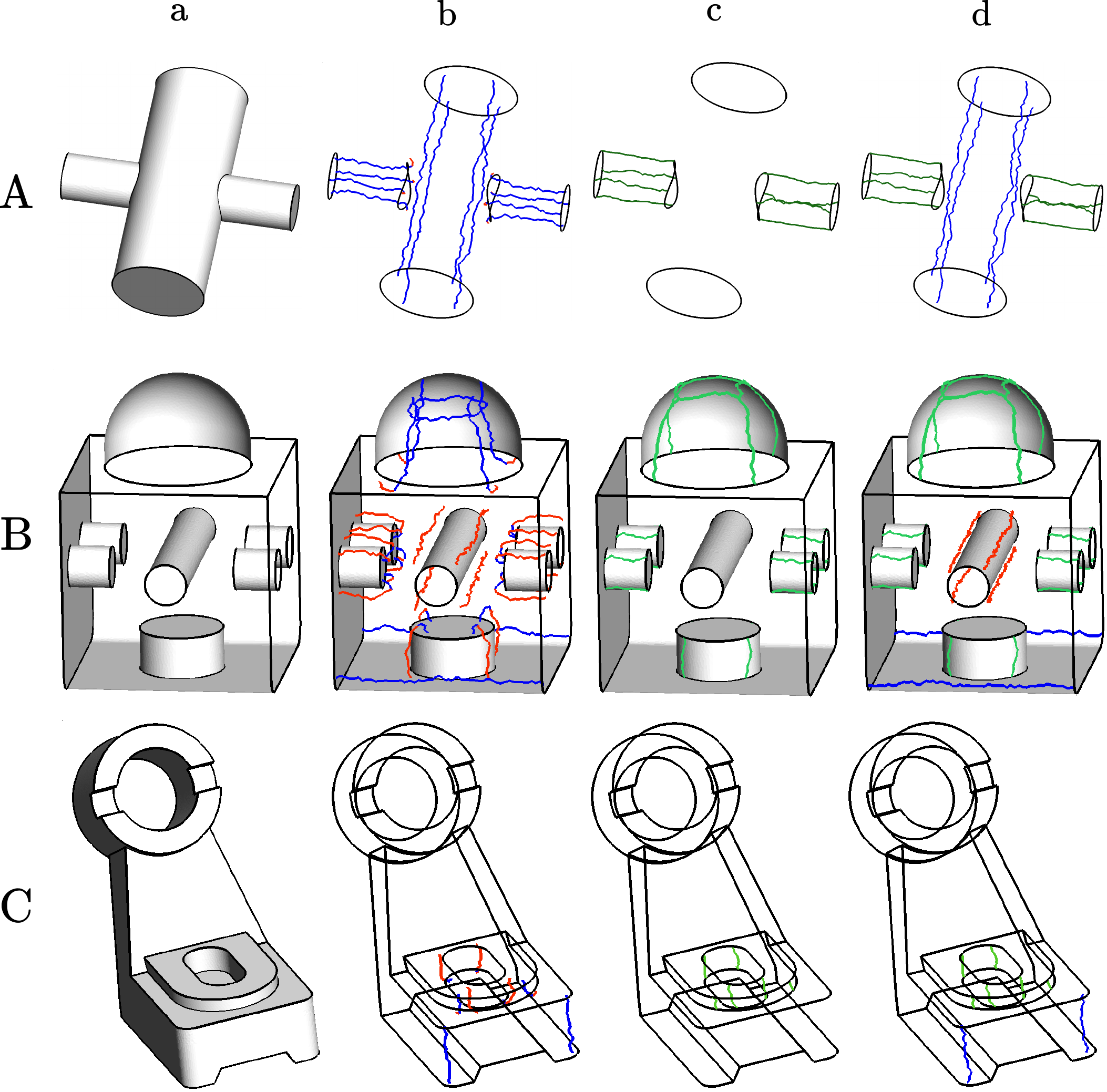} 
    \caption{
        \emph{Model A:} model built from the boolean union of two cylinders.
        \emph{Model B:} model built from boolean operations between cylinders and a sphere, with two fillets
        on the bottom.
        \emph{Model C:} model with various CAD features, from \cite{ledouxIMR}.
        \emph{Col. a:} initial models.
        \emph{Col. b:} singularity graphs of the initial frame fields, which contain many 3-5 singular curves.
        \emph{Col. c:} snapping of the 3-5 singular curves, the snapped curves are shown in dark green.
        \emph{Col. d:} valid singularity graphs of the corrected frame fields, whose boundary conditions have been changed
        according to the snapped curves (in green).
    }
    \label{fig:ex_snap}
\end{figure*}

The singularity snapping correction is applied successfully on the \emph{notch}
model, see Figure \ref{fig:notch_all}.e., on the union of a cube and a
half-sphere, see Figure \ref{fig:hsphere_all}.e. and on three more complicated
CAD models, see Figure \ref{fig:ex_snap}.

On the boxes with imprinted arcs (models A and B on Figure \ref{fig:arcs}), the
3-5 singular curve would be reduced to a single node on the feature curve and
the frame field would be as if there were no imprinted curves.

\paragraph{Geometry and block refinement}

One drawback of this approach is that it produces hexahedral blocks with an
invalid geometry (zero jacobian at some corners), \eg the bottom right block on
Figure \ref{fig:notch_blocks}.b.. We propose to refine the affected blocks in a
post-processing step, as shown on Figure \ref{fig:notch_blocks}.c.. 
For more complex cases, the post-processing refinement can
follow a template-based strategy, such as \cite{schneiders1996}.  To preserve the
topology of a hexahedral mesh, the refinement must be propagated to adjacent blocks
when they share a refined quadrangular face. This is equivalent to sheet insertion.

\paragraph{Limitations}
This approach is only applicable when the 3-5 singular curves are close to the
model boundary, as they are snapped on it. When dealing with CAD models, this
is often the case because the invalid singularities are mostly caused by curved
surface patches, \eg from a boolean operation with a cylinder.

That being said, there are 3-5 singular curves that lie far inside the volume
and whose cause is global, \eg the non-meshable singularity graph of the
\emph{rockerarm} model shown in the Figure 10 of \cite{Liu_2018}. Our
snapping technique does not handle such case.

\section{Conclusions}
To deal with non-meshable 3-5 singular curves, which are induced by CAD feature
curves, we studied four heuristic-based frame field correction strategies.

The \emph{feature curve extrusion} (\S\ref{ss:corr_ic}) and the 
\emph{boundary singular node extrusion} (\S\ref{ss:corr_ssl}) techniques 
try to change globally the frame field topology in the same way a human user
would naively proceed.  This approach does not work when there are interactions
with singularities produced by other features of the model.

A more local and promising approach is to remove the feature curves that induce
the 3-5 singular curves, by transforming them into fillets
(\S\ref{ss:corr_fillet}). It allows additional singular curves that make the
frame field singularity graph valid for hexahedral meshing. However, this
approach is quite impractical because the mapping of the corrected frame field
back to the initial geometry (without fillets) is not straightforward. 

The last approach we explored is to remove the 3-5 singular curves by snapping
them on the model boundary (\S\ref{ss:corr_snap}). The resulting block
decomposition associated to the corrected frame field is similar to the one
obtained by using fillets, minus boundary layers. This technique, which is
also local, is quite efficient and simpler to implement than the \emph{fillet} correction.
However, as the frame field is no longer aligned with the boundary everywhere,
some blocks of the associated decomposition may have flat corners. Fortunately,
these geometric defects can be removed via a post-processing block refinement
procedure.

Short of having a better frame field formulation that do not produce 3-5
singular curves, we recommend using the \emph{singular curve snapping} correction
(\S\ref{ss:corr_snap}), when applicable.

\paragraph{Acknowledgement}
  This research is supported by the European Research Council (project HEXTREME,
  ERC-2015-AdG-694020).

\bibliography{biblio}
\end{document}